\documentclass[twocolumn,aps,prl,reprint,groupedaddress,superscriptaddress]{revtex4}
\usepackage{amsmath}
\usepackage{amssymb}
\usepackage{etoolbox}
\usepackage{graphicx}
\usepackage{color}

\newcommand{\be}{\begin{equation}}
\newcommand{\ee}{\end{equation}}
\def\ba{\begin{aligned}}
\def\ea{\end{aligned}}

\newcommand{\bea}{\begin{eqnarray}}
\newcommand{\eea}{\end{eqnarray}}

\renewcommand{\Im}{{\rm \, Im\,}}

\renewcommand{\hat}[1]{{\widehat #1}}

\renewcommand{\Im}{{\rm Im\,}}

\begin{document}

\title{ Non-ergodic phases in strongly disordered
random regular graphs.}

\author{ B.~L.~Altshuler}
\affiliation{Physics Department, Columbia University, 538 West 120th Street, New York, New York 10027, USA }
\author{E.~Cuevas}
 \affiliation{Departamento de F\'{\i}sica, Universidad de Murcia, E30071 Murcia, Spain}
 \author{ L.~B.~Ioffe}
 \affiliation{CNRS and Universite Paris Sud, UMR 8626, LPTMS, Orsay Cedex, F-91405, France}
\affiliation{L. D. Landau Institute for Theoretical Physics, Chernogolovka, Russia}

\author{V.~E.~Kravtsov}
 \affiliation{Abdus Salam International Center for Theoretical Physics, Strada Costiera 11, 34151 Trieste, Italy}
 \affiliation{L. D. Landau Institute for Theoretical Physics, Chernogolovka, Russia}
 \begin{abstract}

We combine numerical diagonalization with a semi-analytical calculations to prove the existence of the intermediate
non-ergodic but delocalized phase in the Anderson model on disordered hierarchical lattices.
   We suggest a new  generalized  population dynamics
that is able to detect the violation of ergodicity of the delocalized states within the Abou-Chakra,
Anderson and Thouless  recursive scheme.
This result is supplemented by statistics of random wave functions extracted from exact   diagonalization
of the Anderson model on ensemble of disordered Random Regular Graphs (RRG) of N sites with the connectivity $K=2$. By extrapolation
of the results of both approaches to $N\to\infty$
we obtain the fractal dimensions $D_{1}(W)$ and $D_{2}(W)$ as well as the population dynamic exponent $D(W)$
with the accuracy sufficient to claim that they are non-trivial in the broad  interval of disorder strength
$W_{E}<W<W_{c}$. The thorough analysis of the  exact diagonalization results for RRG with $N>10^{5}$ reveals a
singularity in $D_{1,2}(W)$-dependencies which provides a clear evidence for the first order transition between
the two delocalized phases on RRG at $W_{E}\approx 10.0$. We discuss the implications of these results for quantum and classical
non-integrable and many-body systems.

\end{abstract}

\maketitle
 {\it Introduction.}--The concept of Many Body localization (MBL) \cite{BAA} emerged as an attempt to
extend the celebrated ideas of Anderson localization (AL) \cite{Anderson} from one-particle
eigenstates formed by a static random potential to the many-body eigenfunctions of
 macroscopic quantum systems. In Ref.\cite{BAA} it was analytically demonstrated that
cooling of an isolated system of interacting fermions (electrons) with localized
one-particle states leads to a metal - insulator finite-temperature transition
which can be described as MBL in the Fock space. Later on the MBL in various
models (XXZ spin chain subject to a random magnetic field \cite{Ogan-Huse, Ogan-Huse1}, array of
Josephson junctions \cite{LevPino}, etc.) became a subject of intensive theoretical studies.

 The ideas of MBL appear naturally in discussions of
applicability of the conventional Boltzmann-Gibbs statistical mechanics to isolated
many-body systems. This description based on the equipartition postulate
should not be valid for the localized many-body states. Moreover, in Ref. \cite{LevPino} it
was shown that Boltzmann-Gibbs description of the isolated Josephson arrays
most likely remains invalid even in   so called "bad metal" phase where the eigenstates
are extended but not ergodic, e.g. they occupy a vanishing fraction of the Hilbert space.

There are reasons to believe \cite{AGKL} that properties of a one-particle Anderson model
(tight-binding model with on-site disorder) on hierarchical lattices such as the
Bethe lattice (BL) strongly resemble generic properties of a wide class of many-body
models.  BL is characterized by (i) the exponential dependence of the number of sites $N=K^{R}$
(which is analogous to the dimension of the Hilbert space in MBL) on the radius of the tree $R$
with the branching number $K$ and (ii) the absence of loops.
The latter simplifies the problem of AL
as compared to AL in finite dimensions.  In the seminal paper \cite{AbouChacAnd} Abou-Chakra,
Anderson and Thouless    developed an analytical approach to the Anderson model on an infinite
BL that allowed them not only to demonstrate the existence of the AL transition
but also to evaluate the critical disorder with a pretty good accuracy. More recently
some mathematically rigorous results, e.g. the proof of the existence of extended
states and the refined position of the mobility edges were obtained \cite{Aizen, Warzel}.

Nonetheless, the most interesting and the least studied aspect of AL on the BL  is the
statistics of extended wave functions. Recently it has been suggested in Refs.\cite{Biroli,Our-BL}
(see also \cite{Monthus}) that  these statistics may
be multifractal, i.e. extended states in a broad interval of disorder strength may be non-ergodic.
The contradiction of this statement with the earlier studies \cite{MF}  provoked a vigorous
discussion \cite{Biroli15, Rosen, Metz,Tikhonov,Gora}.

Note that the mere formulation of statistics of normalized extended wave functions in a {\it closed} system requires understanding of
the thermodynamic limit of a {\it finite-size} problem.
 For  BL this poses a major problem: a finite fraction of sites  belongs to the boundary  making the results
crucially dependent on the boundary conditions. A known remedy \cite{Biroli,Our-BL} is to consider a Random Regular Graph
(RRG)\cite{radius-graph,MezarParisi},   thus realizing  the boundary-less  hierarchical system  which is locally tree-like.

\begin{figure}[h]
 \center{
  \includegraphics[width=0.9\linewidth]{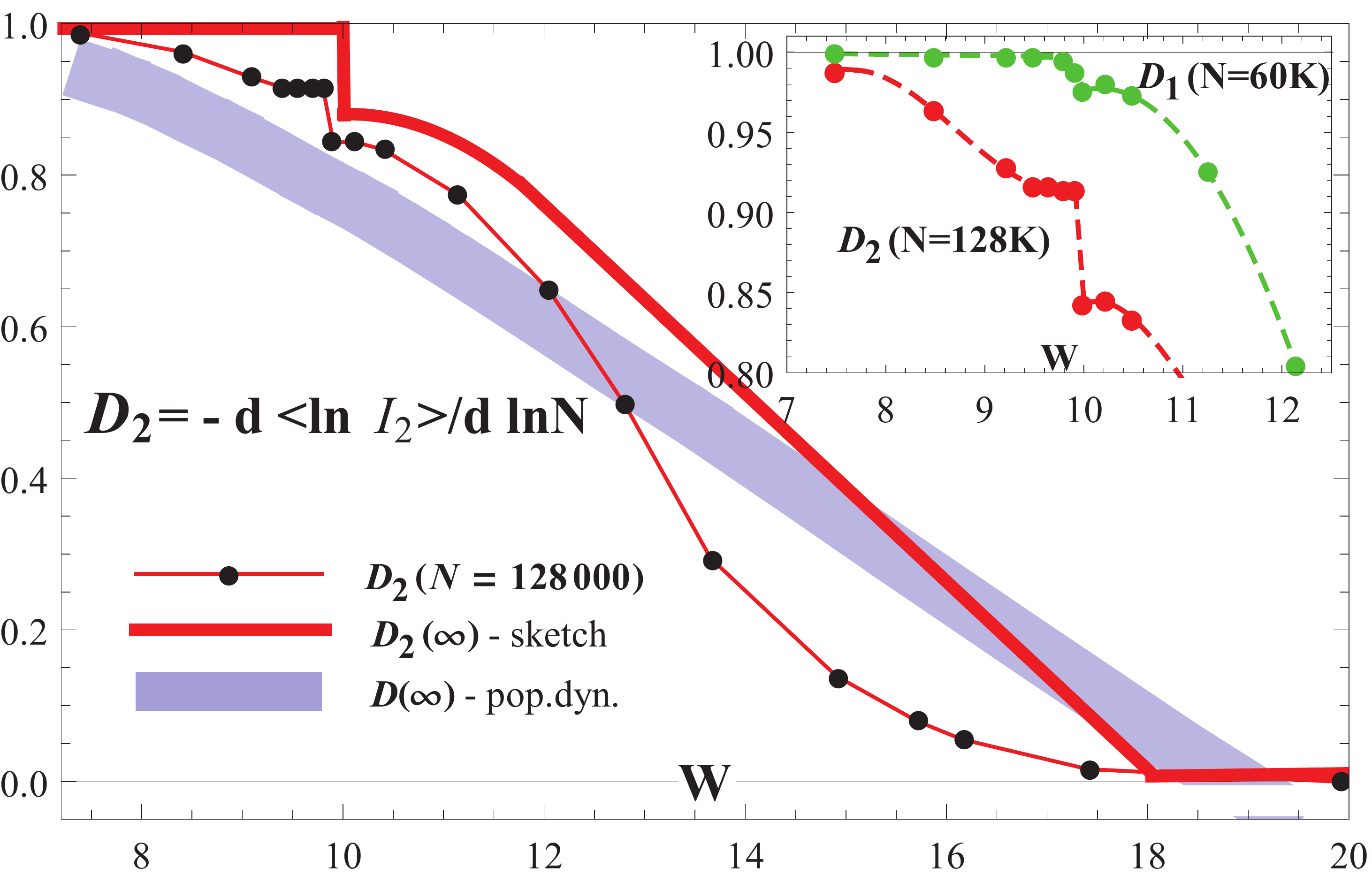}
 }
\caption{(Color online)
Fractal dimensions $D_{2}$ and $D_{1}$ for $K=2$ RRG and the population dynamics exponent $D$
as  functions of disorder strength $W$. The $W$-dependence of  $D$
extrapolated to $N\to\infty$ is presented by
the "brush-painted" blue line which width corresponds to the uncertainty of extrapolation.
In spite of this uncertainty, $D$ is distinctly different from 0 and 1 in a broad interval of $W$ manifesting the non-ergodic (multifractal)
nature of extended wave functions. The red solid line with black data points is a "running" fractal dimension
$D_{2}(N,W)=-d\langle \ln I_{2} \rangle /d \ln N$ obtained by exact diagonalization at the maximal size
$N=128\, 000$ of a disordered RRG. The fat red line is a sketch of the fractal dimension
$D_{2}(N\to\infty,W)\equiv D_{2}(W)$ extrapolated to infinite $N$. Inset: the jump singularity in the "running" fractal dimensions $D_{1}(N=60\,000,W)$ and
$D_{2}(N=128\,000,W)$ manifesting the ergodic transition at $W=W_{E}\approx 10.0$.
}
\label{Fig:sketch}
\end{figure}

In this paper we reformulate the approach of Ref.\cite{AbouChacAnd} in  a way that distinguishes   extended non-ergodic states from ergodic ones.
A new recursive algorithm (similar to population dynamics (PD) \cite{pop-dyn}) of treatment the  Abou-Chakra-Anderson-Thouless (ACAT)
equations \cite{AbouChacAnd}
enables us to
justify semi-analytically the existence of the intermediate extended non-ergodic phase for a BL with $K=2$. Our extensive exact
diagonalization  numerics on the Anderson model on RRG with $N$ up to $128\, 000$ brought up  a strong support for such a phase.
Moreover, we discovered an evidence for the {\it first order transition}  between ergodic (EES) and non-ergodic states (NEES) within the delocalized phase.
Its position  corresponds to the condition for the Lyapunov exponent $L(W,E=0)=\frac{1}{2}\ln K$ discussed in Ref.\cite{Warzel}.
The results are summarized in Fig.\ref{Fig:sketch}.
\\
{\it  The model and fractal dimensions $D_{q}$}--Below we analyze the properties of the eigenfunctions of the Anderson model described by the Hamiltonian:
\begin{gather}\label{And-mod}
\hat{H}=\sum_{i=1}^{N}\varepsilon_{i}\,|i\rangle\langle i|+\sum_{i,j=1}^{N}t_{ij}\,|i\rangle \langle j|.
\end{gather}
Here $\varepsilon_{i}$ are random on-site energies uniformly distributed in the interval $[-W/2, +W/2]$,
 the  connectivity matrix  $t_{ij}$ equals to 1 for nearest neighbors and 0 otherwise. Each site of an infinite BL  has  $K$ neighbors
in the previous generation and
one neighbor in the next generation. In an RRG each site is connected with  $K+1$ randomly chosen other sites. When short loops are neglected RRG becomes
locally equivalent to BL.

Let $|a\rangle$ and $\langle i|a\rangle$ be the normalized eigenstates and wave function coefficients of Hamiltonian Eq.(\ref{And-mod})
in the site representation. One can
introduce the moments   $I_{q}=\sum_{i}|\langle i|a\rangle|^{2q}$ which generically
 scale with the number of the lattice sites $N\gg 1$ as $I_{q}\propto N^{-\tau(q)}$. For localized
states $\tau(q)=0$, while the ergodicity implies $\tau(q)=q-1$. Multifractal non-ergodic
states are characterized by the set of non-trivial fractal dimensions $0<D_{q}=\tau(q)/(q-1)<1$, e.g. $D_{1}=\lim_{q\to 1}D_{q}$
and $D_{2}=\tau(2)$. Exact  diagonalization of a large RRG (see Fig.\ref{Fig:sketch}) suggests that the
fractal dimensions experience a jump from $D_{q}<1$ for $W>W_{E}\approx 10.0$ to $D_{q}=1$
for $W<W_{E}$ manifesting  the first order {\it ergodic transition}.
\\
{\it Generalized recursive algorithm for ACAT equations}--
Following Ref.\cite{AbouChacAnd} we introduce a single-site Green
function, $G^{(k)}_{i}(\omega)=\langle i|(\omega - \tilde{H}_{k})^{-1}|i\rangle$
 for a site $i$ at a generation $k$ of the reduced
Hamiltonian, $\tilde{H}_{k}$ obtained from $\hat{H}$ by disconnecting generations $k$ and $k+1$. The random  Green
functions are  characterized by distribution functions, $P_{k}(G)$.
Individual $G^{(k)}_{i}$ obey the ACAT recursion equation \cite{AbouChacAnd}:
\begin{gather}\label{AChAT}
G^{(k)}_{i}=\frac{1}{\omega-\varepsilon_{i}-\sum_{j(i)}G^{(k-1)}_{j}(\omega)},
\end{gather}
where $j(i)$ are sites at the generation $k-1$ connected to site $i$. These
equations are ill-determined: the pole-like singularities in the right hand sides
 have to be regularized. This is usually achieved by adding an
infinitesimal imaginary part to $\omega\to \omega+i\eta$. The recursion Eq.(\ref{AChAT}) might become
unstable with respect to this addition.
This happens for $W$ below   the
critical disorder of the AL transition $W_{c}$ and manifests the delocalized phase.
For $W > W_{c}$ the solution  $P(G)\propto\delta(\Im G)$ is stable.
The two types of behavior occur generically in a broad class of Anderson
models \cite{Anderson}.

The  spectrum of the Hamiltonian on a finite lattice is given
by a discrete set of energies, $E_{a}$ corresponding to states $|a\rangle$. Although the global
density of states is a sum of delta functions, $\nu(\omega)=\sum_{a}\delta(\omega-E_{a})$, it {\it always} has
a well-defined thermodynamic limit: one introduces an infinitesimal broadening
of each delta function, $\eta$, takes first the limit of the infinite number of sites $N\to\infty$ and
afterwards $\eta\to 0$. As a result, $\nu(\omega)$ tends to a smooth function. In contrast,
for the local density of states (LDoS), $\nu_{i}(\omega)=\sum_{a}|\langle i|a\rangle|^{2}\,\delta(\omega-E_{a})$,
the result of this procedure is
not always  a smooth function. Indeed, in the limit $W\to\infty$ the on-site
states $|i\rangle$ are exact eigenstates and $\nu_{i}(\omega)=\delta(\omega-\varepsilon_{i})$ even for the infinite
system. For finite but large $W$, satellite $\delta$-like peaks appear. The total number
of the peaks is infinite in the thermodynamic limit but almost all of them have
exponentially small weight. Hence the effective number of peaks remains  finite: it
increases as $W$ is decreased and becomes infinite at $W = W_{c}$. At this
point LDoS becomes smooth provided that the limit $N\to\infty$
is taken {\it before} $\eta\to 0$. Note that the opposite order of limits, ($\eta\to 0$ before
$N\to\infty$) always leads to discrete peaks in LDoS.

At $W < W_{c}$ LDoS contains an extensive number $M$ of peaks with
significant weight: $M\to\infty$ as $N\to\infty$. Generally, one expects   $M\propto N^{D}$
with some  $0<D\leq 1$.
For  $\nu_{i}(\omega)$ to be smooth, the broadening $\eta$ should exceed the spacing between the
peaks $\delta_{M}\propto M^{-1}\propto N^{-D}$. Thus, the simultaneous limit $N\to\infty$, $\eta\to 0$, $N^{\gamma}\eta={\rm const}$  results in a
smooth LDoS  {\it iff}
$\gamma<D$.
Studying such generalized limits yields
information on the scaling of the number of peaks, i.e. on the structure of the
eigenfunctions.  Wave functions of {\it ergodic} states are uniformly spread on a lattice, so that $M\propto N$, i.e. $D = 1$ and LDoS is smooth
for any
$\gamma < 1$. We show below that  in a broad interval of disorder strengths in the delocalized regime  $D=D(W) <1$  and equals to the  critical value of $\gamma$
corresponding to the transition between a smooth and a singular LDoS,
$D(W)=
\gamma_{c}(W)$.

For $W < W_{c}$ (delocalized regime) and an infinitesimal  $\eta >0$, $\Im G$ increases exponentially with the number of recursion steps $n$ in Eq.(\ref{AChAT})
describing an {\it infinite} tree:
\begin{gather}\label{exp}
\Im G \propto \eta\,e^{\Lambda n}.
\end{gather}
For a {\it finite} RRG of size $N$,   $n < \ln N/\ln K$ \cite{radius-graph}. For larger $n$ the loops terminate the exponential growth of a typical $\Im G$  limiting it  by
$\Im G \propto \eta\,N^{\Lambda/\ln K}$. Thus  for  $\nu_{i}(\omega)\sim N^{-D}\sum_{a} \eta/[(\omega -E_{a})^{2}+\eta^{2}]\approx
\int dE_{a}\,\eta/[E_{a}^{2}+\eta^{2}]$ to be smooth
(and $\Im G \sim 1$ independent of $\eta$) $\eta$ should  scale as
$\eta\propto N^{-\Lambda/\ln K}$, i.e.
\begin{gather}\label{Lambda-D}
D(W)=\Lambda(W)/\ln K.
\end{gather}

Ideally, one would
deal with infinitely small $\eta\to 0$ in order to determine the exponent
$\Lambda$. However, the limited precision of any numerical computation makes it impossible in practice:
for any realistic initial $\Im G\neq 0 $, the value of $\Im G$ becomes significant after few recursions.
To avoid this problem   we  included an additional
step to the recursion Eq.(\ref{AChAT}):
\begin{gather}\label{new step}
\Im G^{(k)}_{i}\to e^{-\Lambda_{k}}\,\Im G^{(k)}_{i},
\end{gather}
so we keep the {\it typical} imaginary part fixed and
$k$-independent: ${\rm exp}\langle \ln \Im G^{(k)}_{i}\rangle_{k}=\delta$ (where $\langle ...\rangle_{k}$
denotes averaging over all sites $i$ in the $k$-th generation). As soon as the stationary distribution of $G$ is reached in
this recursive procedure, $\Lambda_{k}\to \Lambda$.

To realize  this algorithm we adopted a modified {\it population dynamics} (PD) method \cite{pop-dyn}.
In each step we used the set of $N_{p}$ Green functions
$G^{(k)}_{i}$ ("population")
 obtained at the previous step and new on-site energies $\varepsilon_{i}$ to generate $N_{p}$
new Green functions $G^{(k+1)}_{i}$
 according to Eq.(\ref{AChAT}) in which each site is connected
to $K$ randomly chosen sites of the previous population set.

\begin{figure}[h]
\centering
\includegraphics[width=0.90\linewidth,angle=0]{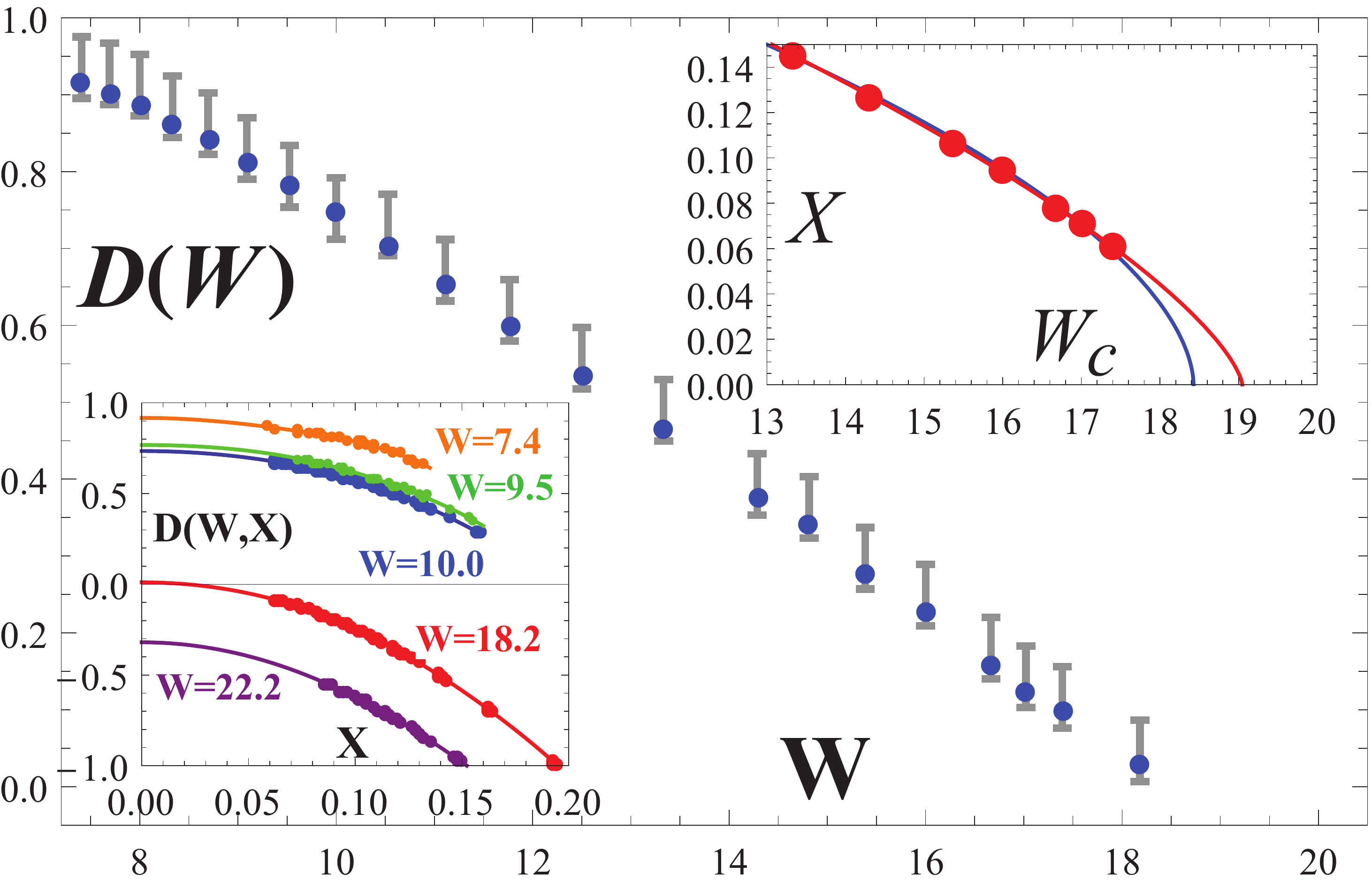}
\caption{(Color online) The population dynamic exponent $D(W)$ (blue points with grey error bars) extrapolated to $N=\infty$ and $\delta=0$ for $K=2$. The condition $D(W_{c})=0$
yields $W_{c}=18.4_{-0.2}^{+0.4}$. In a broad interval of $W<W_{c}$ we obtained $D(W)$ distinctly different from the ergodic limit $D(W)=1$.
{\it Lower inset}: The collapse of data for  a fixed $W$ and different $N_{p}$, $\delta$  to a function $D(W,X)$ of
$X=-1/\ln(N^{-1}_{p}+a\delta^{b})$. Extrapolation to $X=0$ gives  the population dynamic exponent $D(W)$. The delocalized phase corresponds
to $1\geq D(W)>0$, whereas in the localized phase $D(W)<0$.
 {\it Upper inset}: the finite-size critical disorder $W_{c}(X)$ defined as $D(W_{c}(X),X)=0$ and its extrapolation to $X=0$ by the power-law
fit
$W_{c}(X)=W_{c}-a X^{\frac{1}{\nu}}$ with $W_{c}=18.4$, $\nu=0.56$ (blue) and $W_{c}=19.0$, $\nu=0.7$ (red).
Without extrapolation the value of $W_{c}$ at maximal population size $N_{p}^{*}\sim 10^{8}$ is $W_{c}(N_{p}^{*})\approx 17.5$.
  }
 \label{Fig:D-W}
\end{figure}

In order to obtain $D(W)$ one needs to take the limits $N_{p}\to \infty$, $\delta\to 0$ of $D(N_{p},\delta,W)$.
The convergence turns out to be
slow (logarithmic) resulting in a considerable uncertainty in $D(W)$.  Luckily,  $D(N_{p},\delta,W)$ depends
only on the combined variable $X=-1/\ln(N_{p}^{-1}+a \delta^{b})$,
with $a,b\sim 1$, rather than on $\ln N_{p}$ and $\ln \delta$ separately. Extrapolation of $D(W,X)$ to
$X=0$ yields  $D(W)$ shown in Fig.\ref{Fig:D-W}.
The lower inset of Fig.\ref{Fig:D-W} shows the collapse of the data for
several $N$ and $\delta$ from the intervals $10^3 < N < 10^8$ and $10^{-3} < \delta < 10^{-17}$.
Since $b\approx 0.5$,   one needs exceptionally small   $\delta$ to reach
small $X$. This required computation with higher than usual precision.

 Note that the  exponent $\Lambda$ is a property of an infinite BL, $N=\infty$.  Therefore $\Lambda$ is free of the  finite-size effects
which dominate the moments $I_{q}(N)$ at $N<N_{c}$, where
 the correlation volume $N_{c} \sim {\rm exp}[1/\Lambda(W)]$ diverges  at
$W\to W_{c}$.
The uncertainty of extrapolation of $\Lambda$
to $N_{p}\to \infty$ and $\delta\to 0$ turns out to be small enough not to raise doubts
that $0 < D < 1$ at least in the interval  $10<W<18$ for $K = 2$.
Additional support of existence of
the phase with $0<D<1$ comes from the analytical  solution to Eq.(\ref{AChAT}) in the large-$K$ limit \cite{forthcoming}.
It turns out that in this limit:
\be\label{D-large-K}
D(W)=\frac{\ln\left(\frac{W_{c}^{2}}{W^{2}}\right)}{\ln K}.
\ee
It follows from  Eq.(\ref{D-large-K})  that $D(W)=0$ and $D(W)=1$ correspond to the values   $L=\ln K$ and $L=\frac{1}{2}\ln K$ of the Lyapunov exponent $L=\ln W$ discussed in Ref.\cite{Warzel}.

{\it Exact diagonalization on RRG.}-- While ACAT approach is commonly believed   to describe well the localized phase of RRG, its
applicability   in the delocalized  regime requires further inverstigation.
\begin{figure*}[t]
 \centering{
 \includegraphics[width=0.45\linewidth]{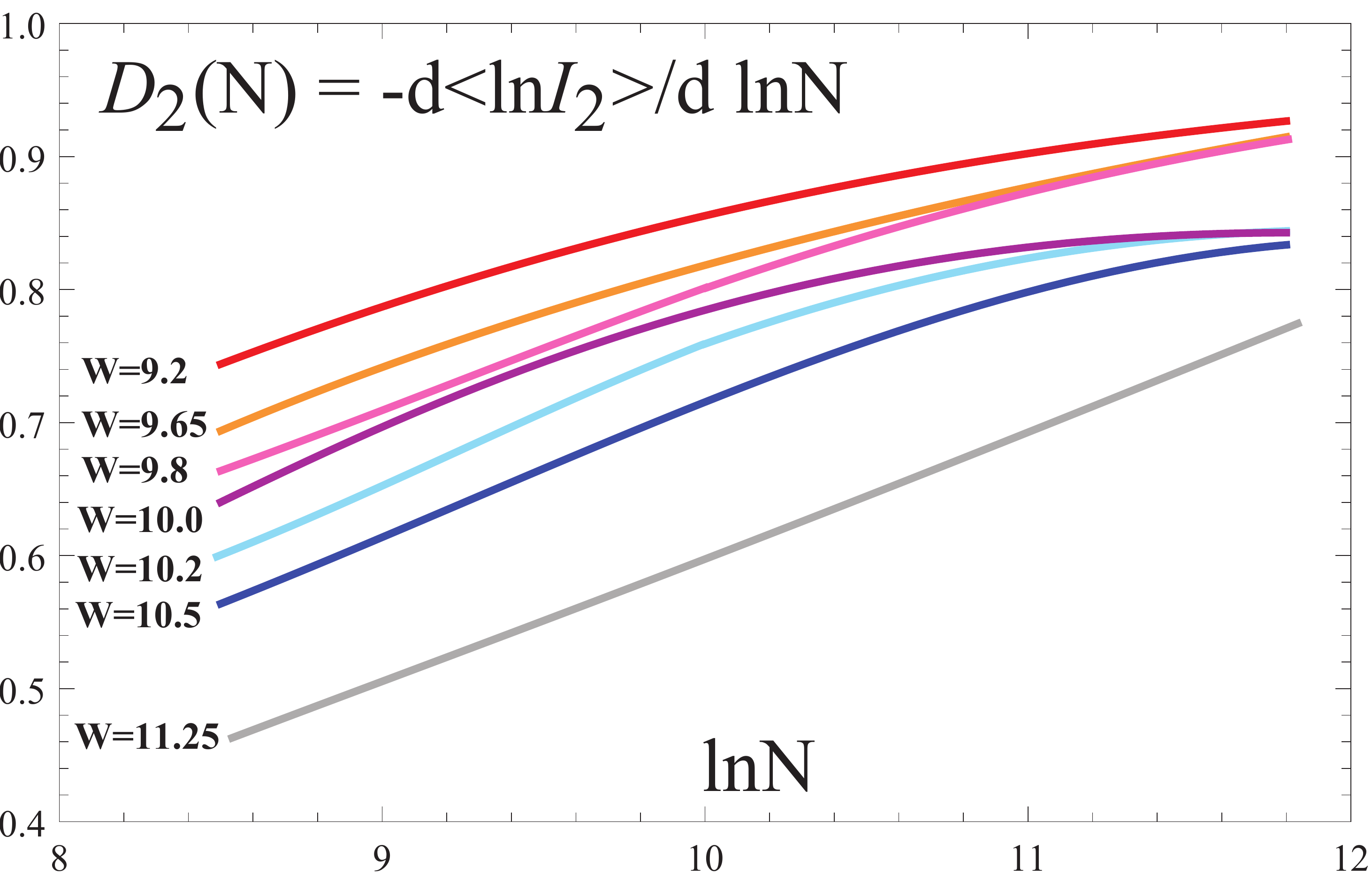}
  \includegraphics[width=0.45\linewidth]{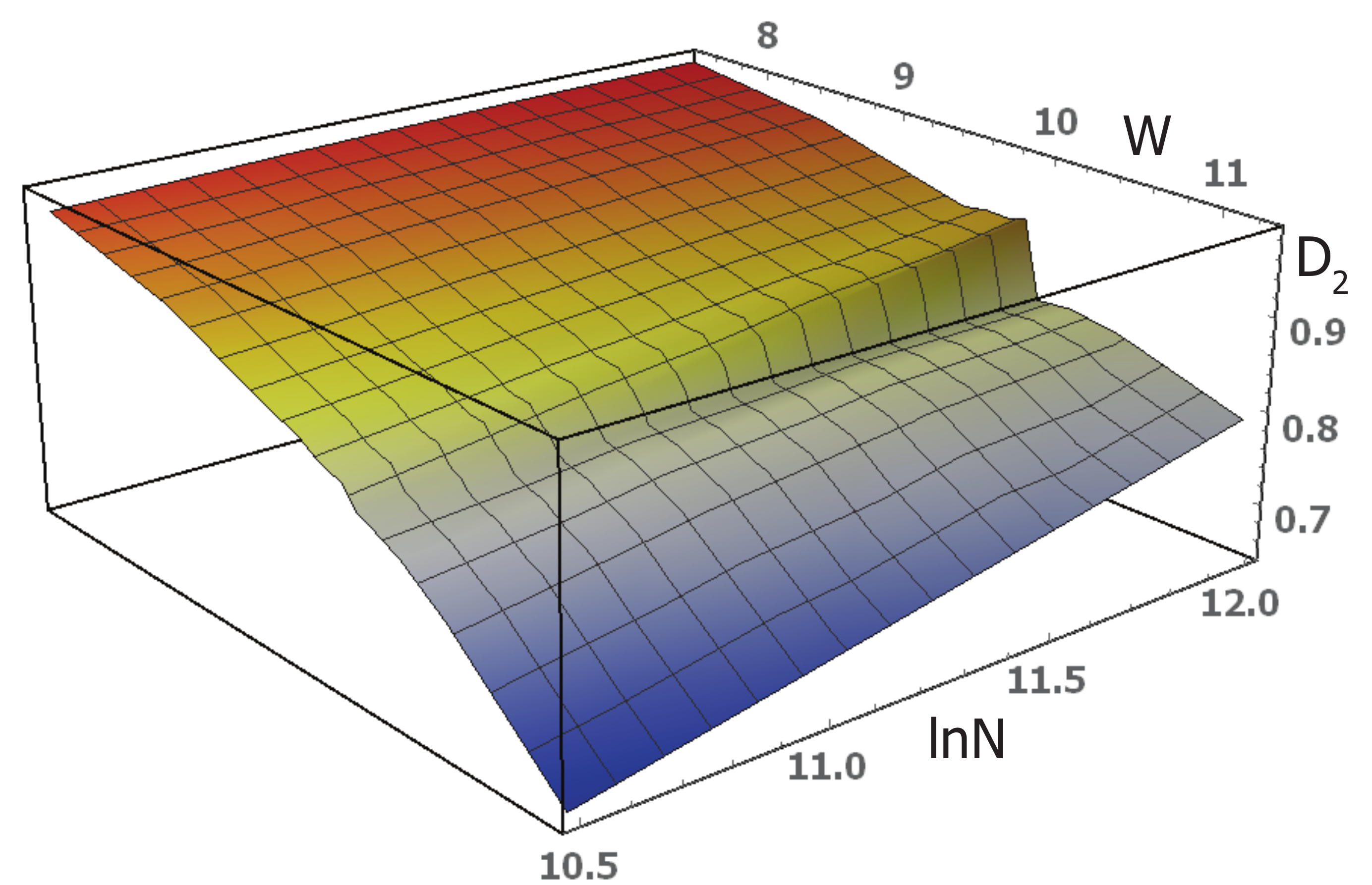}
 }
\caption{(Color online) Left panel: $D_{2}(N,W)$ deep in the delocalized phase.
The curves tend to converge to two different values of $D_{2}$ for $W=W_{E}+0$ and $W_{E}-0$, where $W_{E}\approx 10.0$.
Right panel: Formation of a jump in $D_{2}(W)$.
}
 \label{Fig:moments10}
\end{figure*}
We performed a direct study of the Anderson model on
RRG by exact diagonalization  at the system sizes $N$ up to $128\,000$ in the range of disorder strength $7.5 < W< 20$.
The main focus was on calculating the inverse participation ratio
 $I_{2}=\sum_{i}|\langle i|a\rangle|^{4}$ and the Shannon
 entropy $S= - \sum_{i}|\langle i|a\rangle|^{2}\,\ln(|\langle i|a\rangle|^{2})$
 for the eigenstates $|a\rangle$ with energies $E_{a}$ near the band center.
 The expected asymptotic behavior of the typical averages at $N\to\infty$ is \cite{Our-BL}:
\be\label{asym}
\langle \ln I_{2}\rangle = - D_{2}\,\ln N + c_{2},\;\langle \ln S\rangle = D_{1}\,\ln N +c_{1},
\ee
where $\langle ...\rangle$ are the averages both over the ensemble of RRG with fixed connectivity $K=2$
and over random on-site energies $\varepsilon_{i}$, $D_{1,2}$ are the multifractal dimensions and $c_{1,2}\sim 1$. The derivatives
$D_{2}(N,W)= - d \langle \ln I_{2}\rangle/d\ln N$ and $D_{1}(N,W)=d \langle \ln S\rangle/d \ln N$
should saturate at $D_{2}$ and $D_{1}$, respectively in the limit $N\to \infty$.

\begin{figure}[h]
 \center{
  \includegraphics[width=0.9\linewidth]{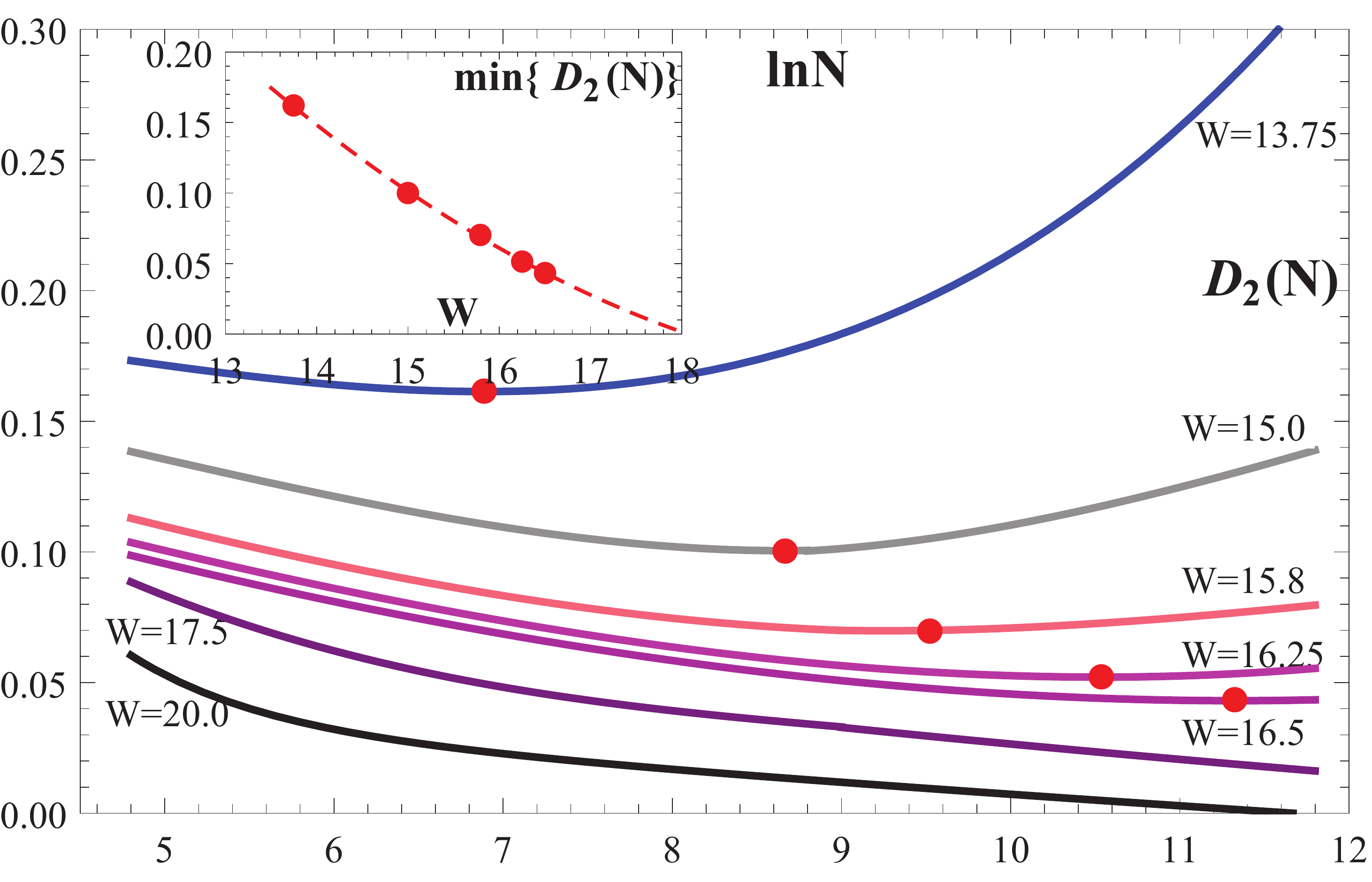}
 }
\caption{(Color online)
 $D_{2}(N,W)$ close to the localization transition at $W=W_{c}$. The $N$-dependence show minima (red spot) for $W<W_{c}$
at $N_{{\rm min}}\to\infty$  as $W\to W_{c}$ \cite{Tikhonov}. Inset: $D_{2}(N_{{\rm min}},W)$ as a function of $W$.
Extrapolation by a second-order polynomial  gives $W_{c}=18.1\pm 0.5$.
}
\label{Fig:moments18}
\end{figure}

 We present the results for $D_{2}(N,W)$ deep in the delocalized phase (Fig.\ref{Fig:moments10}) and close to the localization
 transition (Fig.\ref{Fig:moments18}).
Note two important features on these plots: (i) an abrupt change of behavior for $W$ close to $10$ and (ii)
a minimum in the $N$-dependence of $D_{2}(N,W)$  (recently reported in \cite{Tikhonov}) in the vicinity of AL transition:
as $W\to W_{c}-0$,
$D_{2}(N_{{\rm min}},W)$ at the minimum and $1/\ln N_{{\rm min}}$ vanish. Extrapolation of
$D_{2}(N_{{\rm min}},W)$ leads to $W_{c}=18.1\pm 0.5$ (see inset to Fig.\ref{Fig:moments18}) in agreement with
 PD results, Fig.\ref{Fig:D-W}.

A striking  result of the exact diagonalization is the existence of a jump in both $D_{2}(N,W)$ and $D_{1}(N,W)$ shown
in Fig.\ref{Fig:sketch}. A feature, which is almost invisible at small $N$
evolves to a more and more abrupt jump as $N$ increases above $60\,000$ (see Fig. \ref{Fig:moments10}, right panel).
Extrapolation of $D_{2}(N,W)$ to $N\to\infty$  for $W<10.0$ gives
$D_{2}=D_{2}(N\to \infty,W)=1$, whereas $D_{2}(W=10.0)=0.86\pm 0.02$. We conclude that on RRG at $W=W_{E}\approx 10.0$ there is
a first order transition  from the non-ergodic delocalized phase at $W>W_{E}$ to the ergodic one at
$W<W_{E}$.

{\it Conclusion.}
The existence of the non-ergodic phase of the BL Anderson model together with the similarity of this model with generic many-body
ones gives basis for far-reaching speculations. The point is that in contrast to the conventional Anderson localization, which is
the property of any wave dynamics, the MBL is a genuine quantum phenomenon. Indeed, in the classical limit,
a weakly perturbed integrable system with $d > 2$ degrees of freedom   always demonstrates some diffusion in the phase space known as
Arnold diffusion\cite{Arnold}. Although the celebrated Kolmogorov Arnold Moser (KAM) theorem \cite{KAM}
guarantees the survival of the vast majority of the invariant tori the chaotic part of the phase space is
connected (unless $d = 2$), thus allowing the diffusion for arbitrary weak perturbation.
 Therefore one should not expect MBL in the classical limit. On the other hand the glassy
states of matter without doubts exist for any $\hbar$  including $\hbar=0$ and are obviously not ergodic. It is safe to assume that the extended non-ergodic phase of the MBL
models is not qualitatively different from a classical glassy state \cite{Mezard-glass}. Therefore our arguments in
favor of the existence of the delocalized non-ergodic phase of the BL Anderson model and the true phase transition
between the ergodic and non-ergodic states can  be considered as arguments in favor of glassy states being distinct
states of matter and their transition to fluids being a true phase transition.
\\
{\it Acknowledgement} We appreciate hospitality at KITP of University of California at Santa Barbara under the NSF grant
No.NSF PHY11-25915, at KITPC in Beijing (V. E. K.) and at the Center for Theoretical Physics of Complex Systems (Daejeon, S.Korea)   where the research has been carried out. E.C. thanks partial financial support by the Murcia Regional Agency of Science
and Technology (project 19907/GERM/15). The research of L.B.I. was partially supported  by the Russian Science
Foundation grant No. 14-42-00044. We are grateful to G. Biroli, E. Bogomolny, J. T. Chalker,
  M. V. Feigelman, M. Foster, I. M. Khaymovich,  G. Parisi, V. Ros, A. Scardicchio, M. A. Skvortsov, V. N.  Smelyanskiy, K. S. Tikhonov and S. Warzel
for stimulating discussions.

\begin{widetext}
\section{Supplementary material}

\subsection{Analysis of statistics of inverse participation ratio for RRG and 3D Anderson model in the delocalized phase.}
\begin{figure}[h]
\center{\includegraphics[width=0.6\linewidth]{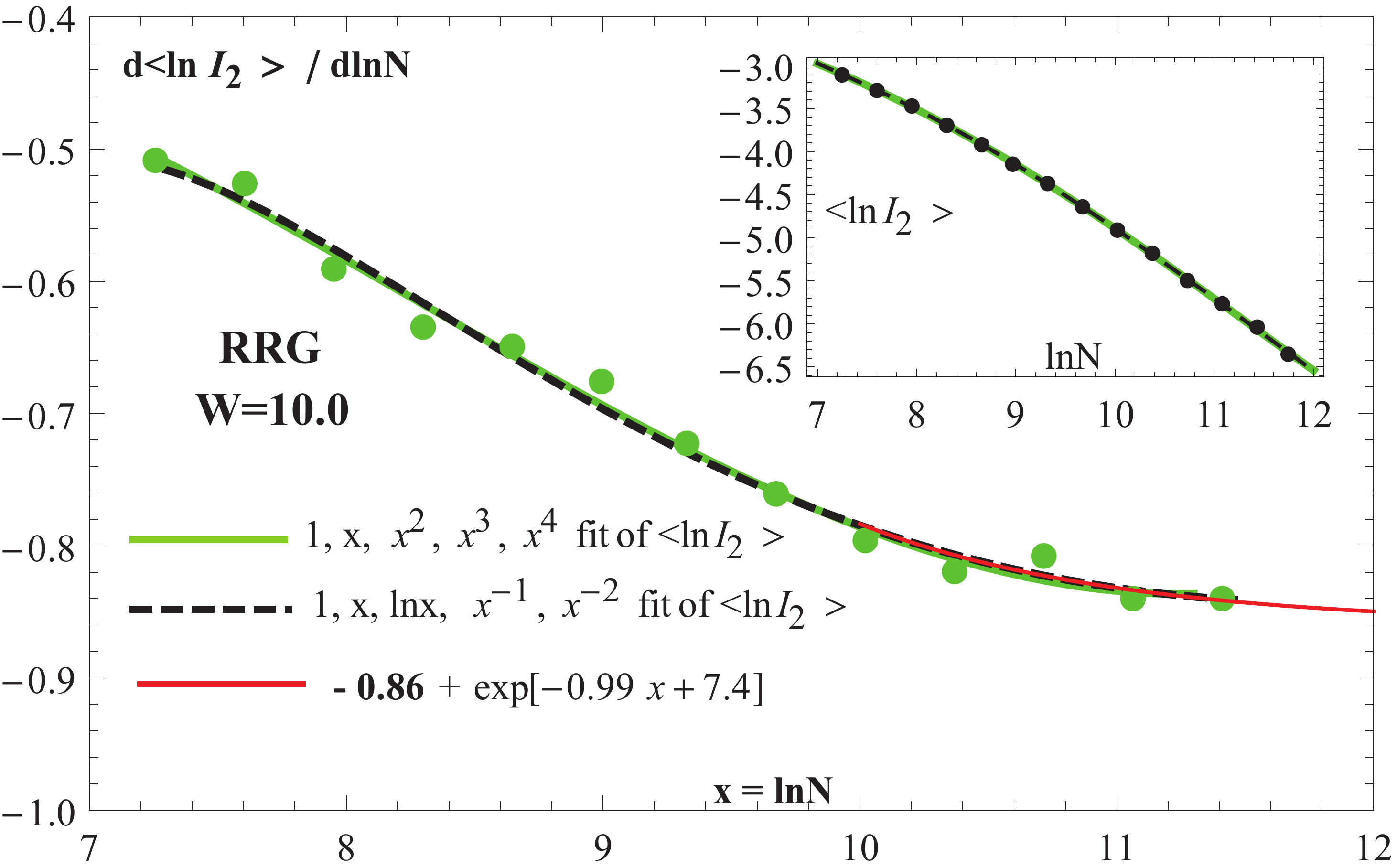}}
\caption{(Color online) The running fractal dimension   $-D_{2}(N,W)$ for RRG at $W=10.0$ as a function of $x=\ln N$. Main plot: The solid green and dashed black lines are the derivatives of the two interpolating functions shown in the inset. The green spots are the discrete derivatives obtained using the raw data points. The red solid line is the best power law in $N$ fit of the large-$N$ tail of the (almost coinciding) green and black dashed curves. It yields $D_{2}=0.86\pm 0.02$ distinctly different from the ergodic limit $D_{2}=1.0$. Inset: the raw data points for the ensemble averaged $\langle \ln I_{2}\rangle$ and the two different interpolating functions (green and dashed black curves).} \label{Fig:RRG10}
\end{figure}
In order to evaluate the   fractal dimensions $D_{1}(W)$ and $D_{2}(W)$ we performed an exact diagonalization of matrix Hamiltonians drawn
from the  ensemble of (i) different RRG with branching number $K=2$ and a fixed total number of sites $N$ ranging from $1000$ to $128\,000$ and
(ii) different realizations ($\sim 1000$ for the largest $N=96\,000$ and $N=128\,000$) of the independent random on-site energies with
the box distribution characterized by the disorder strength $W$. For the largest $N=96\,000$ and $N=128\,000$ we analyzed $\sim 50$ eigenstates with
energies close to the band-center.

For each value of $W$ this step of the  calculation results in a set of $N$-dependent ensemble averages
\be
\langle \ln I_{2}\rangle=\left \langle\ln \sum_{i} |\langle a|i\rangle|^{4} \right\rangle,\;\;\;\langle \ln S\rangle=\left \langle\ln \sum_{i} |\langle a|i\rangle|^{2} \ln(1/ |\langle a|i\rangle|^{2}) \right\rangle,
\ee
 where $\langle a|i\rangle$ are the wave function coefficients of a state $a$ at a site $i$. The averages $\langle \ln I_{2} \rangle$ are presented by points in the inserts of  Fig.\ref{Fig:RRG10}.-Fig.\ref{Fig:AM10}.
We define the running fractal dimensions $D_{2}(N,W)$ and $D_{1}(N,W)$ as  derivatives
\be
D_{2}(N,W)=-d\langle \ln I_{2}\rangle/d\ln N,\;\;\;D_{1}(N,W)=d \langle \ln S\rangle/d \ln N,
\ee
 respectively, since the $N\to \infty$ limits of these derivatives are by definition $D_{2}(W)$ and $D_{1}(W)$. It is convenient to first find a smooth multi-parameter approximations of the dependencies  of $\langle \ln I_{2}\rangle$ and $\langle\ln S \rangle$ on $\ln N$ and then evaluate the derivatives of these smooth interpolating functions. Typical 5-parameter interpolating functions are shown in Fig.\ref{Fig:RRG10}.
We also evaluated  discrete  derivatives (green points in Fig.\ref{Fig:RRG10}-Fig.\ref{Fig:AM10}).

The last step was to find the best power-law (exponential in $x=\ln N$) fit
\be
D_{1,2}(N,W)=D_{1,2}(W)-c/N^{\gamma} = D_{1,2}(W) - {\rm exp}[-\gamma x + {\rm cnst}].
\ee
 of the large-$N$ {\it tail} of the smooth interpolating functions $D_{1,2}(N,W)$ (red solid curves in Fig.\ref{Fig:RRG10}.-Fig.\ref{Fig:AM10}).
Our values of fractal dimensions $D_{1,2}(W)=D_{1,2}(\infty,W)$ follow from this extrapolation.
The procedure described above yields $D_{1,2}(W)$ not distinguishable form 1
for $W<10.0$ (e.g. for $W=9.5$ shown in Fig.\ref{Fig:RRG9.5} it gives $D_{2}(W)=1.0\pm 0.02$). At the same time for $W=10.0$ the best
fit results in $D_{2}(10.0)=0.86\pm 0.02$ distinctly different from 1. We checked that the analogous procedure for the three-dimensional
Anderson model in the delocalized phase with $W=10.0$ yields $D_{2}=1.0\pm 0.01$ (Fig.\ref{Fig:AM10}) with all the dependencies similar
to the ones for RRG at $W=9.5$ ( Fig.\ref{Fig:RRG9.5}). This analysis led us to the conclusion that the finite-$N$ jump in $D_{1,2}(N,W)$
shown in Fig.1 of the main paper survives the thermodynamic limit $N\to \infty$ and implies the first order transition to the ergodic phase
for $W<W_{E}\approx 10.0$.
\begin{figure}[h]
\center{\includegraphics[width=0.6\linewidth]{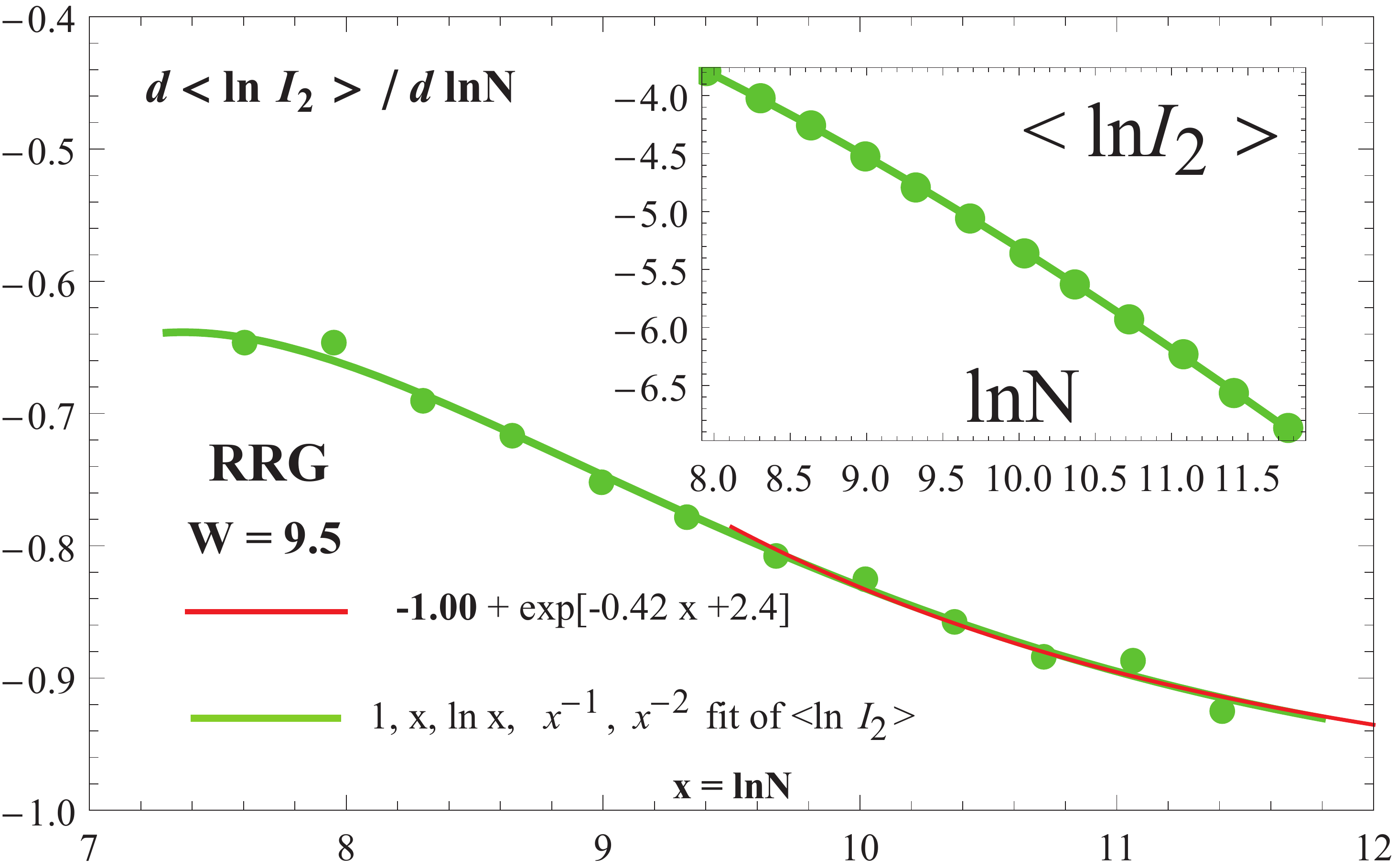}}
\caption{(Color online)The running fractal dimension  $-D_{2}(N,W)$ for RRG at $W=9.5$. Main plot: the solid green line is the
derivative of the smooth 5-parameter interpolating function for the ensemble averaged $\langle \ln I_{2}\rangle$ shown in the inset.
The red line is the best power-law in $N$ fit of the large-$N$ tail of the solid green line. It gives $D_{2}=1.0\pm 0.02$.} \label{Fig:RRG9.5}
\end{figure}
\begin{figure}[h]
\center{\includegraphics[width=0.6\linewidth]{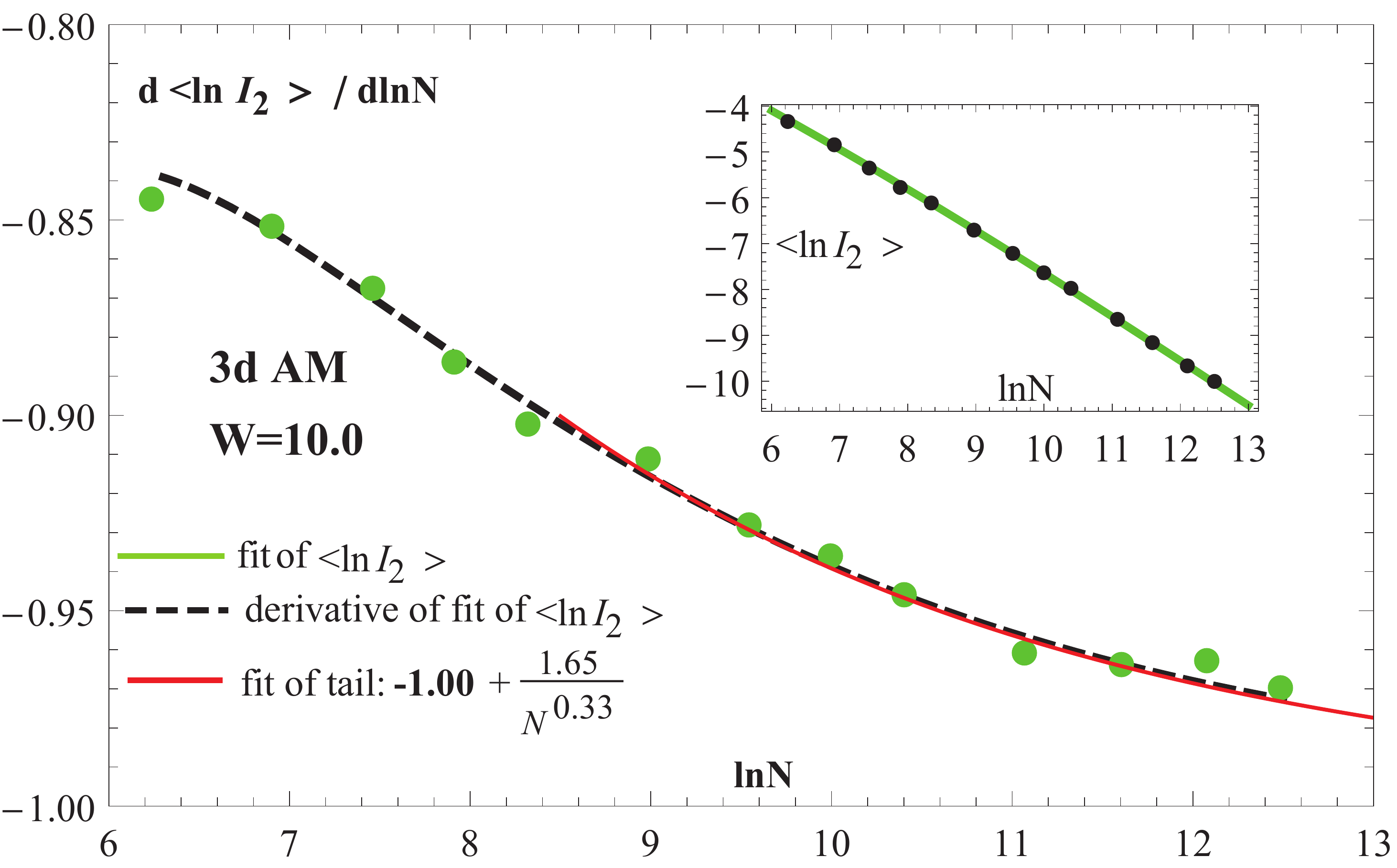}}
\caption{(Color online) The running fractal dimension  $-D_{2}(N,W)$ for the three-dimensional Anderson model ($W_{c}\approx 16.5$)
in the delocalized phase at $W=10.0$. The solid green line in the inset is the 5-parameter interpolating function for the black raw data
points for the ensemble averaged $\langle \ln I_{2}\rangle$. The black dashed line in the main plot is the  derivative of the smooth
interpolating function. The red solid line is the best power-law in $N$ fit to the large-$N$ tail of the black dashed line. It gives
$D_{2}=1.0\pm 0.01$ similar to the RRG with $W<10.0$. } \label{Fig:AM10}
\end{figure}
Another evidence of the transition at $W=W_{E}\approx 10.0$ is the singularity at $W=W_{E}$ in the speed of evolution
with $\ln N$ of $D_{2}(N,W)$. In Fig.\ref{Fig:speed} it is shown that while $D_{2}(N,W)$ generally increases with
increasing $\ln N$, the speed of this evolution reaches a deep minimum at $W=W_{E}\approx 10.0$, making the convergence better
at $W=10.0$ than both for $W>10.0$ and for $W<10.0$.
\begin{figure}[h]
\center{\includegraphics[width=0.6\linewidth]{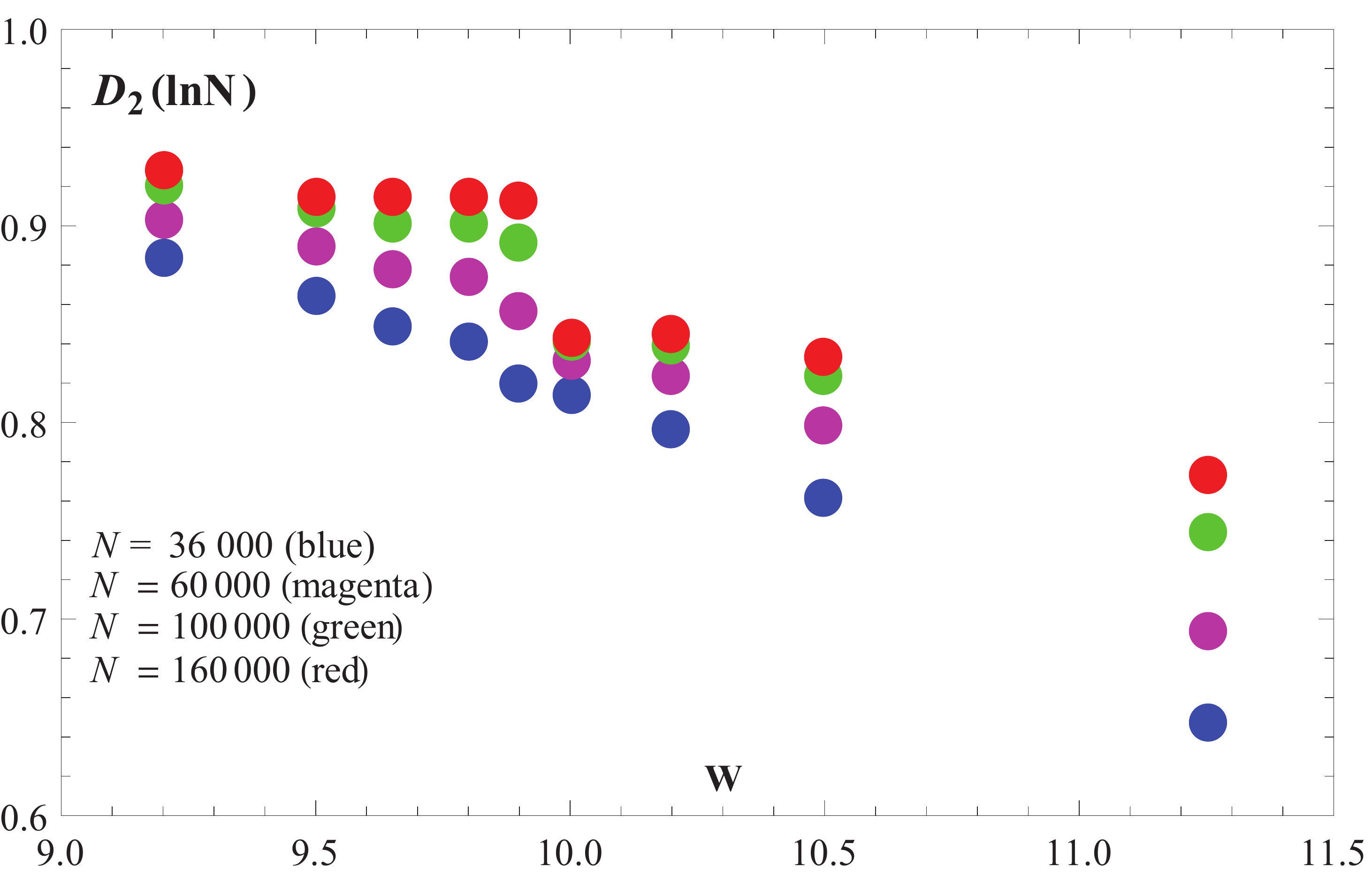}}
\caption{(Color online) The speed of evolution of $D_{2}(N,W)$ with $N$. Blue, magenta, green and red spots correspond to
$N=36\,000,60\,000,100\,000,160\,000$, respectively, in the "stroboscopic shot" of the evolution. The speed of evolution has a sharp minimum at $W=10$. }
\label{Fig:speed}
\end{figure}
In Fig.\ref{Fig:curvature} we further illustrate this point:  the difference $D_{2}(N=128\,000,W)-D_{2}(N=13\,000,W)$ shows a sharp singularity at $W=W_{E}$. We
would like to emphasize that no extrapolation was employed.
\begin{figure}[h]
\center{\includegraphics[width=0.6\linewidth]{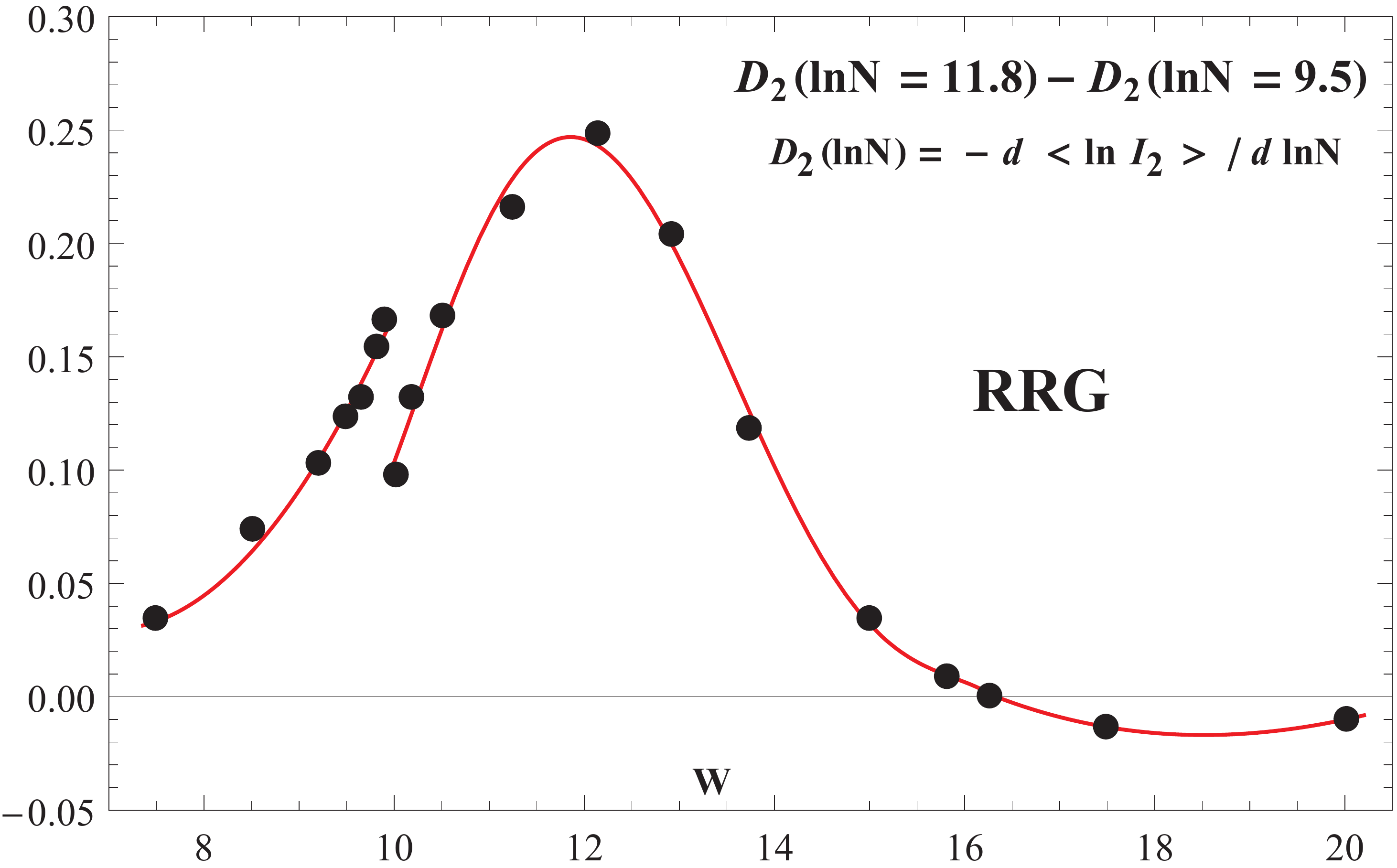}}
\caption{(Color online)  The difference $D_{2}(N=128\,000,W)-D_{2}(N=13\,000,W)$. It shows a sharp singularity at $W=W_{E}\approx 10.0$. }
\label{Fig:curvature}
\end{figure}

Both features, (i) the jump in $D_{1,2}(W)$, and (ii) the anomaly in the speed of evolution, prove that the phase
that exists on RRG for $W>10.0$ may not be ergodic.
We argue that   the singularity of $D_{2}(W)$ function  at $W=W_{E} \approx 10.0$  completely excludes the
possibility of the states in the interval $W_{E} < W < W_{c}$ being ergodic.
\end{widetext}
\end{document}